\documentclass[aps,twocolumn,showpacs]{revtex4-1}
\usepackage{amssymb}
\usepackage{amsmath}
\usepackage{amscd}
\usepackage{graphicx}
\usepackage{subfigure}
\usepackage{float}
\usepackage{array}
\begin{document}

\title[Short Title]{Ground state of three qubits coupled to a harmonic oscillator with ultrastrong coupling}

\author{Li-Tuo Shen}
\author{Zhen-Biao Yang}
\email{zbyang@fzu.edu.cn}
\author{Rong-Xin Chen}

\affiliation{Lab of Quantum Optics, Department of Physics, Fuzhou
University, Fuzhou 350002, China}

\begin{abstract}
We study the Rabi model composed of three qubits coupled to a
harmonic oscillator without involving the rotating-wave
approximation. We show that the ground state of the three-qubit Rabi
model can be analytically treated by using the transformation
method, and the transformed ground state agrees well with the
exactly numerical simulation under a wide range of qubit-oscillator
coupling strengths for different detunings. We use the pairwise
entanglement to characterize the ground state entanglement between
any two qubits and show that it has an approximately quadratic
dependence on the qubit-oscillator coupling strength. Interestingly,
we find that there is no qubit-qubit entanglement for the ground
state if the qubit-oscillator coupling strength is large enough.
\end{abstract}

\maketitle

\noindent

\section{Introduction}

Recent experimental progresses related to the qubit-oscillator
system in ultrastrong coupling regime have been reported in
different light-matter interaction systems
\cite{PRB-78-180502-2008,PRB-79-201303-2009,Nature-458-178-2009,Nature-6-772-2010,
PRL-105-237001-2010,PRL-105-196402-2010,PRL-106-196405-2011,
Science-335-1323-2012,PRL-108-163601-2012,PRB-86-045408-2012}, where
the coupling strength between a single qubit and a single oscillator
reaches a significant fraction of the oscillator and qubit
frequencies. In this ultrastrong coupling regime, the ubiquitous
Jaynes-Cummings model \cite{IEEE-51-89-1963} under the rotating-wave
approximation (RWA) is expected to break down leading to a mass of
unexplored physics and giving rise to fascinating quantum phenomena
\cite{NJP-13-073002-2011,PRL-109-193602-2012,PRA-87-013826-2013,
PRA-59-4589-1999,PRA-62-033807-2000,PRB-72-195410-2005,PRA-74-033811-2006,
PRA-77-053808-2008,PRA-82-022119-2010,PRL-107-190402-2011,PRL-108-180401-2012,PRA-87-022124-2013,
PRA-86-014303-2012}. For example, superradiance transition
\cite{PRA-87-013826-2013}, vacuum Rabi-splitting
\cite{PRB-78-180502-2008,NJP-13-073002-2011}, photon blockade
\cite{PRL-109-193602-2012}, Bloch-Siegert shift
\cite{PRL-105-237001-2010}, and plasmonic effect
\cite{PRB-86-045408-2012}.

Since the Hamiltonian of a qubit-oscillator system contains
counter-rotating wave terms that make the computational subspace
unclosed, the fully analytical solution to the ground state of this
Hamiltonian in the ultrastrong coupling limit is still not found.
Although the spectrum and eigenfunction of the Rabi model beyond the
RWA are known by numerical diagonalization in a truncated
finite-dimensional Hilbert space
\cite{JPA-29-4035-1996,EPL-96-14003-2011}, the analytical solution
to the qubit-oscillator system beyond the RWA is necessary for
clearly capturing the fundamental physics. Such an analytical
treatment has the potential to be extended to more complicated
models for the implementation of the quantum information processing
(QIP) \cite{PRA-81-042311-2010}. Therefore, various mathematics
approaches have been proposed to analytically obtain the ground
state properties of the single-qubit Rabi model in the ultrastrong
coupling regime \cite{RPB-40-11326-1989,PRB-42-6704-1990,
PRL-99-173601-2007,EPL-86-54003-2009,PRA-80-033846-2009,
PRL-105-263603-2010,PRA-82-025802-2010,PRL-107-100401-2011,
EPJD-66-1-2012,PRA-86-015803-2012,PRA-85-043815-2012,
PRA-86-023822-2012}. For example, the generalized-RWA mehtod
\cite{PRL-99-173601-2007,EPL-86-54003-2009} functions well when the
qubit frequency is smaller than the oscillator frequency, the
variational treatment
\cite{PRA-82-025802-2010,RPB-40-11326-1989,PRB-42-6704-1990}
reasonably captures the properties of ground state in the
single-qubit Rabi system which is very hard to be generalized to
multi-qubit Rabi systems, and the transformation method is a
perturbation expansion and has been successfully applied to the
single-qubit Rabi system
\cite{PRA-86-015803-2012,EPJB-38-559-2004,PRB-75-054302-2007,EPJD-59-473-2010}.

Recently, the Tavis-Cummings model beyond the RWA has been extended
to the multi-qubit case by using an adiabatic approximation method
when the qubit frequency is far larger than the oscillator frequency
\cite{PRA-85-043815-2012}, and the ground state of the nearly
resonant Rabi model of two qubits coupled to a harmonic oscillator
has been analytically treated by using both the variational and the
transformation methods \cite{arXiv-1303-3367v2-2013}. The Rabi model
of three and more qubits coupled to a common harmonic oscillator in
the ultrastrong coupling regime has more potential applications in
QIP \cite{PRL-107-190402-2011,PRL-108-180401-2012} than that the
single-qubit Rabi model has, such as protected quantum computation
\cite{PRL-107-190402-2011}, which is expected to be very promising
with the circuit QED architecture.

However, the ground states of the three- and more-qubit Rabi models
in the ultrastrong coupling regime have not been extensively
studied. Recently, Braak has generalized the method based on the
$Z_{2}$ symmetry \cite{PRL-107-100401-2011} to the three-qubit Dicke
model \cite{arXiv-1304-2529v1-2013} to analytically determine the
system's spectrum, which is dependent on the composite
transcendental function defined through its power series. However,
this method can not be extended to determine the concrete form of
the ground state.

Different from the Ref. \cite{arXiv-1304-2529v1-2013}, we focus here
on the analytic ground state of the three-qubit Rabi model in the
ultrastrong coupling regime by the transformation method. By mapping
the three-qubit Rabi model into a solvable Jaynes-Cummings-like
model, we show that the ground state energy and the ground state of
this three-qubit Rabi model can be approximately determined by the
analytic expression based on the transformation method, which agrees
well with the exactly numerical simulation in the ultrastrong
coupling regime under different detunings. The ground state
entanglement between any two qubits is characterized by using the
pairwise entanglement and has a quadratic dependence on the
qubit-oscillator coupling strength, which can be approximately
determined within a wide range of parameters. The interesting
feature in the ground state entanglement exists in its maximum
value, which decreases quickly to zero and never increases again as
the qubit-oscillator coupling strength is large enough.

\section{Transformed ground state}

The Hamiltonian of three identical qubits coupled to a harmonic
oscillator without the rotating-wave approximation is ($\hbar=1$)
\begin{eqnarray}\label{e1}
H&=&\frac{1}{2}w_{a}(J_{+}+J_{-})+w_{c}a^{\dagger}a+g(a^{\dagger}+a)J_{z},
\end{eqnarray}
where $a$ and $a^{\dagger}$ are respectively the annihilation and
creation operators of the harmonic oscillator with frequency
$w_{c}$. $J_{l}$ $\{ l=\pm,z \}$ describes the collective atomic
operator of a spin-$\frac{3}{2}$ system, satisfying the angular
momentum commutation relations $[J_{z},J_{\pm}]=\pm J_{\pm}$ and
$[J_{+},J_{-}]=2J_{z}$. Physically, the spin-$\frac{3}{2}$ system is
nontrivial and the states are entangled in terms of individual qubit
configurations. $w_{a}$ is the transition frequency of each qubit.
$g$ represents the collective qubit-oscillator coupling strength.

The key point in this paper is to determine the ground state energy
$E_{g}$ and the ground state $|\phi_{g}\rangle$ for the three-qubit
Rabi system in the ultrastrong coupling regime, where
$H|\phi_{g}\rangle=E_{g}|\phi_{g}\rangle$. To derive the analytic
ground state, we define $|m\rangle_{a}$ to be an eigenvector of
$J_{z}$, i.e., $J_{z}|m\rangle_{a}=m|m\rangle_{a}$
($m=-\frac{3}{2},-\frac{1}{2},\frac{1}{2},\frac{3}{2}$). Besides, we
will respectively use $|X\rangle_{f}$ and $|0\rangle_{f}$ to
represent the coherent field state with the real amplitude $X$ and
the vacuum field state.

In what follows we extend the transformation method used in the
single- and two-qubit Rabi models
\cite{EPJD-59-473-2010,arXiv-1303-3367v2-2013} to the three-qubit
Rabi model. To transform the Hamiltonian $H$ into a mathematical
form without the counter-rotating wave terms, we apply a unitary
transformation to the Hamiltonian $H$:
\begin{eqnarray}\label{e2}
H^{'}&=&e^{S}He^{-S},
\end{eqnarray}
with
\begin{eqnarray}\label{e3}
S&=&\chi(a^{\dagger}-a)J_{z},
\end{eqnarray}
where $\chi$ is a variable to be determined. Then the transformed
Hamiltonian $H^{'}$ is decomposed into three parts
\cite{arXiv-1303-3367v2-2013}:
\begin{eqnarray}\label{e4}
H^{'}&=&H_{0}^{'}+H_{1}^{'}+H_{2}^{'},
\end{eqnarray}
with
\begin{eqnarray}\label{e5-e7}
H_{0}^{'}&=&\eta w_{a}
J_{x} -(2g\chi-w_{c}\chi^2)J_{z}^2+w_c a^{\dagger}a,\\
H_{1}^{'}&=&(g-w_{c}\chi)(a^{\dagger}+a)J_{z}+i\eta
w_{a}\chi(a^{\dagger}-a)J_{y}, \\ H_{2}^{'}&=&w_{a}J_{x}\{
\cosh\big[\chi(a^{\dagger}-a)\big]-\eta \}\cr&& +iw_{a}J_{y}\bigg\{
\sinh\big[ \chi(a^{\dagger}-a)\big] -\eta\chi(a^{\dagger}-a)
\bigg\},
\end{eqnarray}
where
$\eta=$$_{f}$$\langle0|\cosh[\chi(a^{\dagger}-a)]|0\rangle_{f}=\exp[-\chi^2/2]$.
The terms $\cosh\big[\chi(a^{\dagger}-a)\big]$ and
$\sinh\big[\chi(a^{\dagger}-a)\big]$ in $H_{2}^{'}$ have the
dominating expansions:
\begin{eqnarray}\label{e8-e9}
\cosh[ \chi(a^{\dagger}-a) ]&=&\eta+O(\chi^2),\\
\sinh[\chi(a^{\dagger}-a) ]&=&\chi\eta(a^{\dagger}-a)+O(\chi^3),
\end{eqnarray}
here $O(\chi^2)$ and $O(\chi^3)$ represent the double- and
multi-photon transition processes containing higher-order operators
for $a^{\dagger}$ and $a$, which can be neglected as an
approximation when $\chi$ is small. Note that the transformation
method works well only if $\chi g/(w_{a} +w_{c}) << 1$, and it fails
in the ultrastrong coupling regime where $\chi > 1$. Thus,
$H^{'}\simeq H_{0}^{'}+H_{1}^{'}$. By now, our approximation
procedure is the same as that in Ref. \cite{arXiv-1303-3367v2-2013}.
However, the main difference exists in the diagonalization for
$H^{'}_{0}$ with the collective qubit vectors. Such an operator
$H_{J}^{'}=\eta w_{a}J_{x}-(2g\chi-w_c\chi^2)J_{z}^{2}$ appearing in
the qubit basis $\Gamma_{a}=\{ |-\frac{3}{2}\rangle_{a},
|-\frac{1}{2}\rangle_{a}, |\frac{1}{2}\rangle_{a},
|\frac{3}{2}\rangle_{a}\}$ represents a renormalized four-level
qubit system. This is different from the two-qubit Rabi system
\cite{arXiv-1303-3367v2-2013} which is a renormalized three-level
atomic system corresponding to the diagonalization for $H^{'}_{0}$.
Through diagonalizing the operator $H_{J}^{'}$ in the qubit basis
$\Gamma_{a}$, we obtain the renormalized qubit eigenvector
$|\varphi_{k}\rangle_a$ with the eigenvalue $\lambda_{k}$
($k=1,2,3,4$) as following:
\begin{eqnarray}\label{e10}
\lambda_{1}&=&5A-\frac{1}{2}B-2\sqrt{4A^2+AB+\frac{1}{4}B^2},\cr
|\varphi_{1}\rangle_a&=&\frac{1}{N_{1}}\bigg(-|-\frac{3}{2}\rangle_{a}+K_{1}|-\frac{1}{2}\rangle_{a}-K_{1}|\frac{1}{2}\rangle_{a}+|\frac{3}{2}\rangle_{a}\bigg),\cr
\lambda_{2}&=&5A+\frac{1}{2}B-2\sqrt{4A^2-AB+\frac{1}{4}B^2},\cr
|\varphi_{2}\rangle_a&=&\frac{1}{N_{2}}\bigg(|-\frac{3}{2}\rangle_{a}-K_{2}|-\frac{1}{2}\rangle_{a}-K_{2}|\frac{1}{2}\rangle_{a}+|\frac{3}{2}\rangle_{a}\bigg),\cr
\lambda_{3}&=&5A-\frac{1}{2}B+2\sqrt{4A^2+AB+\frac{1}{4}B^2},\cr
|\varphi_{3}\rangle_a&=&\frac{1}{N_{3}}\bigg(-|-\frac{3}{2}\rangle_{a}+K_{3}|-\frac{1}{2}\rangle_{a}-K_{3}|\frac{1}{2}\rangle_{a}+|\frac{3}{2}\rangle_{a}\bigg),\cr
\lambda_{4}&=&5A+\frac{1}{2}B+2\sqrt{4A^2-AB+\frac{1}{4}B^2},\cr
|\varphi_{4}\rangle_a&=&\frac{1}{N_{4}}\bigg(|-\frac{3}{2}\rangle_{a}-K_{4}|-\frac{1}{2}\rangle_{a}-K_{4}|\frac{1}{2}\rangle_{a}+|\frac{3}{2}\rangle_{a}\bigg),\cr&&
\end{eqnarray}
and
\begin{eqnarray}\label{e11}
A&=&-\frac{1}{4}(2g\chi-w_c\chi^2),\cr B&=&\eta w_a,\cr
K_{1}&=&\frac{1}{\sqrt{3}B}\bigg(8A+B+4\sqrt{4A^2+AB+\frac{1}{4}B^2}\bigg),\cr
K_{2}&=&\frac{1}{\sqrt{3}B}\bigg(8A-B+4\sqrt{4A^2-AB+\frac{1}{4}B^2}\bigg),\cr
K_{3}&=&\frac{1}{\sqrt{3}B}\bigg(8A+B-4\sqrt{4A^2+AB+\frac{1}{4}B^2}\bigg),\cr
K_{4}&=&\frac{1}{\sqrt{3}B}\bigg(8A-B-4\sqrt{4A^2-AB+\frac{1}{4}B^2}\bigg),
\end{eqnarray}
where $N_{k}=\sqrt{2+2K_{k}^2}$ $(k=1,2,3,4)$ is the normalization
factor for the eigenvector $|\varphi_{k}\rangle_{a}$.
\begin{figure}
\center
  \includegraphics[width=1\columnwidth]{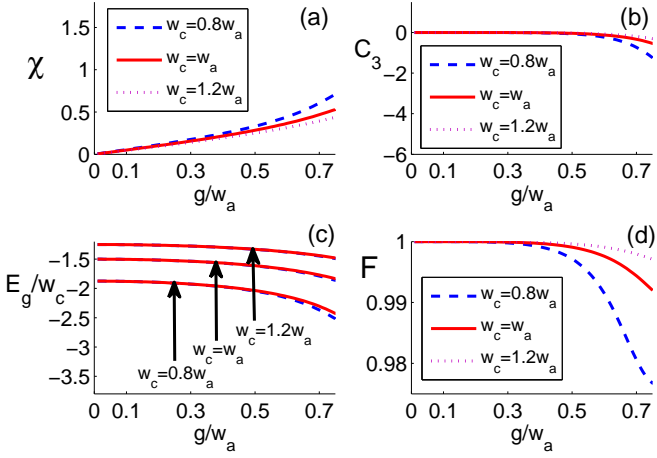} \caption{(Color
  online) (a) Numerical solutions for the variable $\chi$ that makes
  $C_{1}=0$. (b) The $C_{3}$ value as a function of $g$ for different qubit-oscillator detunings.
   (c) The ground state energy as a function of the coupling
  strength $g$. The solid line represents the transformed ground state energy
  $E_{g}^{'}$ and the dash line represents the exact
  ground state energy $E_{g}$. (d) The fidelity $F$ of the ground state
  $|\phi_{g}^{'}\rangle$ obtained by the transformation method.
  }\label{Fig.1.}
\end{figure}
For $\chi g\approx g\approx w_{a}$, the eigenvalues here are
arranged in the decreasing order through the numerical simulation,
i.e., $\lambda_{1}<\lambda_{2}<\lambda_{3}<\lambda_{4}$. Therefore,
$H^{'}$ can be expanded with the above renormalized eigenvectors:
\begin{eqnarray}\label{e12}
H^{'}&&=
\sum_{k=1}^{4}\lambda_{k}|\varphi_{k}\rangle_{a}\langle\varphi_{k}|+
\bigg[(C_{1}a+C_{2}a^{\dagger})|\varphi_{1}\rangle_{a}\langle\varphi_{2}|\cr&&
+(C_{3}a+C_{4}a^{\dagger})|\varphi_{1}\rangle_{a}\langle\varphi_{4}|
+(C_{5}a+C_{6}a^{\dagger})|\varphi_{2}\rangle_{a}\langle\varphi_{3}|\cr&&
+(C_{7}a+C_{8}a^{\dagger})|\varphi_{3}\rangle_{a}\langle\varphi_{4}|+H.c.\bigg]+w_c
a^{\dagger}a,
\end{eqnarray}
where $C_{x}(x=1,2,3,...,8)$ is the coefficient depending on the
variable $\chi$. It is obvious to see that $C_{1},C_{3},C_{5}$ and
$C_{7}$ represent the coupling strengths of the corresponding
counter-rotating wave terms with respect to the renormalized
eigenvectors in Eq. (\ref{e12}).

Similar to the single- and two-qubit Rabi systems
\cite{EPJD-59-473-2010,arXiv-1303-3367v2-2013}, the main task after
transforming the Hamiltonian $H$ into $H^{'}$ is to eliminate the
counter-rotating wave terms for the eigenvector with the lowest
eigenenergy. The major obstacle here is to remove the different
coupling coefficients $C_{1}$ and $C_{3}$ of two counter-rotating
wave terms for the eigenvector $|\varphi_{1}\rangle_{a}$
simultaneously. This is very different from the single-qubit
\cite{EPJD-59-473-2010} and the two-qubit Rabi models
\cite{arXiv-1303-3367v2-2013}, which just have one counter-rotating
wave term for the approximate ground state vector. Although it is
not possible to simultaneously remove the coefficients $C_{1}$ and
$C_{3}$ for all the values of $\chi$, we find that the conditions
$C_{1}=0$ and $C_{3}\approx0$ can be both satisfied when
$0\leq\chi\leq0.5$, meaning two counter-rotating wave terms for the
approximate ground state vector $|\varphi_{1}\rangle_{a}$ can both
be eliminated if the qubit-oscillator interaction is not too strong.
The coefficients $C_{1}$ and $C_{3}$ have the following analytical
forms:
\begin{eqnarray}\label{e13-e14}
C_{1}&=& (3+K_1K_2)(g-w_c\chi)\cr&&-\eta w_a\chi\bigg(
\sqrt{3}K_{1}+2K_{1}K_{2}-\sqrt{3}K_{2}\bigg),\\
C_{3}&=& (3+K_1K_4)(g-w_c\chi)\cr&&-\eta w_a\chi\bigg(
\sqrt{3}K_{1}+2K_{1}K_{4}-\sqrt{3}K_{4}\bigg).
\end{eqnarray}
Therefore, $|\varphi_{1}\rangle_{a}|0\rangle_{f}$ is expected to be
the approximate ground state vector if the conditions $C_{1}=0$ and
$C_{3}\approx0$ are both satisfied, then the ground state
$|\phi_{g}\rangle$ of this three-qubit Rabi system approximates the
transformed ground state $|\phi_{g}^{'}\rangle$:
\begin{eqnarray}\label{e15}
|\phi_{g}^{'}\rangle&=&e^{-S}|\varphi_{1}\rangle_{a}|0\rangle_{f}\cr
&=&\frac{1}{N_{1}}\bigg(-|-\frac{3}{2}\rangle_{a}|\frac{3}{2}\chi\rangle_{f}+K_{1}|-\frac{1}{2}\rangle_{a}|\frac{1}{2}\chi\rangle_{f}\cr&&
-K_{1}|\frac{1}{2}\rangle_{a}|-\frac{1}{2}\chi\rangle_{f}+|\frac{3}{2}\rangle_{a}|-\frac{3}{2}\chi\rangle_{f}\bigg),
\end{eqnarray}
and the transformed ground state energy $E_{g}^{'}$ is:
\begin{eqnarray}\label{e16}
E_{g}^{'}&\simeq&\lambda_{1}\cr
 &=&\frac{5}{4}w_{c}\chi^2-\frac{5}{2}\chi
g-\frac{1}{2}w_{a}e^{-\frac{\chi^2}{2}}-
\bigg[(2g\chi-w_c\chi^2)^2\cr&&-w_{a}(2g\chi
-w_c\chi^2)e^{-\frac{\chi^2}{2}}+w_{a}^{2}e^{-\chi^2}\bigg]^{\frac{1}{2}}.
\end{eqnarray}

According to the condition $C_{1}=0$, the numerical solution of
$\chi$ is plotted as a function of the coupling strength $g$ for
different qubit-oscillator detunings in Fig. 1(a). We find $\chi$
has a proportional relation with $g$: $\chi\simeq\frac{g}{w_a+w_c}$.
By substituting the result $\chi$ from Fig. 1(a) into Eq. (14), we
obtain the corresponding solution for $C_{3}$ in Fig. 1(b), which
shows the conditions $C_{1}=0$ and $C_{3}\approx0$ can be both
satisfied when $0\leq g\leq0.5w_{a}$. This guarantees two
counter-rotating wave terms in the eigenvector
$|\varphi_{1}\rangle_{a}$ are both eliminated if the
qubit-oscillator coupling is not too strong. In Fig. 1(c), we have
estimated the accuracy between the transformed ground state energy
$E_{g}^{'}$ and the exact ground state energy $E_{g}$ under
different qubit-oscillator detunings, in which $E_{g}^{'}$ has an
approximately quadratic dependence on $g$:
\begin{eqnarray}\label{e17}
E_{g}^{'}&\simeq&-\frac{3}{2}w_a-\frac{3}{2w_c+3w_a}g^2.
\end{eqnarray}
We see that the transformed ground state energy achieves the nearly
perfect matching with the exactly numerical value within the
ultrastrong coupling regime $g\leq 0.5w_{a}$. When $g=0.5w_{a}$, the
errors for the transformed ground state energy at $w_{c}=0.8w_{a}$,
$w_{c}=w_{a}$, and $w_{c}=1.2w_{a}$ are $0.49\%$, $0.19\%$, and
$0.07\%$, respectively. Especially, when there is a positive
detuning $w_{c}-w_{a}>0$, the transformed ground state energy fits
much better with the exact value for a wide range of $g$ than that
with the negative qubit-oscillator detuning or the exact
qubit-oscillator resonance, and its error is only $0.9\%$ even when
$g=0.8w_a$ for $w_{c}=1.2w_{a}$. This result coincides with the
variational behavior of $C_{3}$ in Fig. 1(b), in which $C_{3}$ grows
much slower with $w_{c}=1.2w_{a}$ than the case with
$w_{c}=0.8w_{a}$ or $w_{c}=w_{a}$ when the coupling strength
satisfies $g>0.5w_{a}$.

To examine the reliability of the transformed ground state
$|\phi_{g}{'}\rangle$, we use the fidelity $F$, which is defined as
$F=\langle\phi_{g}^{'}|\phi_{g}\rangle$, as a measurement for the
transformation method. From the result of Fig. 1(d), we find that
the exact ground state for the three-qubit Rabi model can be
approximately represented by $|\phi_{g}^{'}\rangle$ within the
ultrastrong coupling regime $0\leq g\leq0.5w_a$. For example, we
obtain the fidelity with high value $F>99\%$ for $g\leq0.5w_a$ under
different qubit-oscillator detunings.
\section{Ground state entanglement}
\begin{figure} \center
  \includegraphics[width=1\columnwidth]{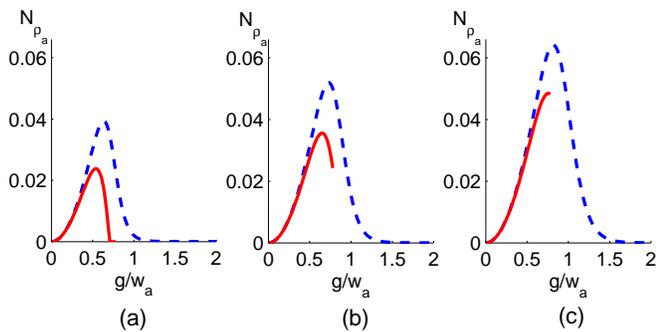} \caption{(Color
  online) The pairwise entanglement $N_{\rho_a}$ as a function of the
  coupling strength $g$ under different qubit-oscillator detunings: (a)
  $w_{c}=0.8w_{a}$; (b) $w_{c}=w_{a}$; (c) $w_{c}=1.2w_{a}$. The
  red solid curves (blue dashed curves) correspond to the pairwise entanglement
  of the transformed (exact) ground state. The blue dashed curves
  vanish at (a) $g/w_{a}=1.22$, (b) $g/w_{a}=1.45$, and (c)
  $g/w_{a}=1.82$. The photon number cutoff we used here is $N_{tr}=30$.
  }\label{Fig.1.}
\end{figure}

To investigate the qubit-qubit entanglement for the present
three-qubit Rabi model in the ground state, in which the
prescription set out for symmetric Dicke states is used, we proceed
to consider the pairwise entanglement
\cite{PRA-68-012101-2003,EPJD-18-385-2002} between any two qubits.

Taking the transformed ground state $|\phi_{g}^{'}\rangle$ in Eq.
(\ref{e15}), the reduced density matrix $\rho_{a}$ of any two qubits
can be written as:
\begin{eqnarray}\label{e18}
\rho_{a}&=&\left(\begin{array}{cccc}
\rho_{11} & 0 & 0 & \rho_{14} \\
0 & \rho_{22} & \rho_{23} & 0 \\
0 & \rho_{32} & \rho_{33} & 0 \\
\rho_{41} & 0 & 0 & \rho_{44}
\end{array}
\right),
\end{eqnarray}
where
\begin{eqnarray}\label{e19}
\rho_{11}&=&\rho_{44}=\frac{N^2-2N+4\langle
J_{z}^{2}\rangle}{4N(N-1)}
=\frac{1}{6}+\frac{1}{3(1+K_{1}^{2})},\cr\cr
\rho_{14}&=&\rho_{41}=\frac{\langle J_{+}^{2}\rangle}{N(N-1)}
=\frac{\sqrt{3}K_{1}}{3(1+K_{1}^{2})}e^{-2\chi^2},
\cr\cr\rho_{22}&=&\rho_{23}=\rho_{32}=\rho_{33}=\frac{N^2-4\langle
J_{z}^{2}\rangle}{4N(N-1)}=\frac{K_{1}^{2}}{3(1+K_{1}^{2})},\cr&&
\end{eqnarray}
and the standard basis in $\rho_{a}$ is $\{$
$|e_{l}\rangle|e_{m}\rangle$, $|e_{l}\rangle|s_{m}\rangle,$
$|s_{l}\rangle|e_{m}\rangle,$ $|s_{l}\rangle|s_{m}\rangle$ $\}$,
with $|e_{l}\rangle$ ($|e_{m}\rangle$) and $|s_{l}\rangle$
($|s_{m}\rangle$) ($l,m=1,2,3;$ and $l \neq m$) denoting the excited
and ground state of the $l$th ($m$th) qubit, respectively.
Therefore, the pairwise entanglement $N_{\rho_{a}}$ can be expressed
as:
\begin{eqnarray}\label{e20}
N_{\rho_{a}}&=&2\max\bigg\{ 0,|\rho_{23}|-\sqrt{\rho_{11}\rho_{44}},
|\rho_{14}|-\sqrt{\rho_{22}\rho_{33}}\bigg\}.\cr&&
\end{eqnarray}
In the ultrastrong coupling regime $g\leq 0.5w_{a}$, we can
numerically verify:
\begin{eqnarray}\label{e21}
N_{\rho_{a}}&=&\frac{2(\sqrt{3}K_{1}e^{-2\chi^2}-K_{1}^2)}{3(1+K_{1}^{2})}\simeq
\frac{1}{4(w_{a}+w_{c})^2}g^2.
\end{eqnarray}
Fig. 2 illustrates the pairwise entanglement $N_{\rho_{a}}$ obtained
from the transformed and the exact ground states versus the coupling
strength $g$ under different detunings. We see that the pairwise
entanglement has a quadratic dependence on $g$ at small coupling
strength, which is mathematically captured by the approximate power
law between $N_{\rho_a}$ and $g$ in Eq. (\ref{e20}). If
$g>0.5w_{a}$, discrepancies for the numerical results between the
transformed and exact ground states become bigger as the coupling
strength increases further. The maximal pairwise entanglement
between any two qubits is determined by the detuning
$\Delta=w_c-w_a$, and the maximal pairwise entanglement in the qubit
system increases with $\Delta$. Interestingly, the pairwise
entanglement $N_{\rho_{a}}$ decreases quickly to zero after reaching
its maximum, and remains at zero even when the coupling strength $g$
increases further, which means that there is no qubit-qubit
entanglement in the ground state of such a model any more if the
coupling strength $g$ is large enough. For example, the pairwise
entanglement decreases to zero for $g=1.5w_a$ at the exact resonant
case and never increases again even when $g$ increases further. This
feature is distinguished from the result of Ref.
\cite{arXiv-1303-3367v2-2013} and can be explained as follows. When
the field and one qubit is traced out, the first and fourth terms of
Eq. (15) do not result in the entanglement of the other two qubits.
In other words, the pairwise entanglement is contributed by the
$W$-state components $|-\frac{1}{2}\rangle_ {a}$ and
$|\frac{1}{2}\rangle_ {a}$. The coefficients of the terms involving
$|-\frac{1}{2}\rangle_ {a}$ and $|\frac{1}{2}\rangle_ {a}$ quickly
drop to zero when $g$ is large enough, resulting in the vanishing of
pairwise entanglement.


\section{Conclusion}

In summary, we have shown that the ground state of the three-qubit
Rabi model in the ultrastrong coupling regime can be approximately
treated by the transformation method. The transformed ground state
fits very well with the exact ground state for different detunings
even when the coupling strength $g$ increases to $0.5w_a$. When
$g=0.5w_a$, the error of the transformed ground state energy is only
$0.19\%$ at $w_a=w_c$, and the fidelity for the transformed ground
state keeps higher than $99\%$ when $g\leq0.5w_{a}$ under different
qubit-oscillator detunings. Finally, we use the pairwise
entanglement to analytically examine the qubit-qubit entanglement,
and the result shows that the ground state entanglement has an
approximately quadratic dependence on the qubit-oscillator coupling.
Interestingly, we find that there is no ground state entanglement if
the qubit-oscillator coupling strength is large enough.

\section{Acknowledgement}

This work is supported by the Major State Basic Research Development
Program of China under Grant No. 2012CB921601, the National Natural
Science Foundation of China (NSFC) under Grant No. 11247283, and
funds from Fuzhou University under Grant No. 022513, Grant No.
022408, and Grant No. 600891.



\begin{thebibliography}{99}




\bibitem{PRB-78-180502-2008} A.~A.~Abdumalikov Jr, O.~Astafiev,
Y.~Nakamura, Y.~A.~Pashkin, and J.~S.~Tsai, Phys. Rev. B 78,
180502(R) (2008).
\bibitem{PRB-79-201303-2009} A.~A.~Anappara, S.~DeLiberato,
A.~Tredicucci, C.~Ciuti, G.~Biasiol, L.~Sorba, and F.~Beltram, Phys.
Rev. B 79, 201303(R) (2009).
\bibitem{Nature-458-178-2009} G.~G\"{u}nter, A.~A.~Anappara,
J.~Hees, \emph{et al.}, Nature 458, 178 (2009).
\bibitem{Nature-6-772-2010} T.~Niemczyk, F.~Deppe, H.~Huebl,
 \emph{et al.}, Nature 6, 772 (2010).
\bibitem{PRL-105-237001-2010} P.~Forn-D\'{\i}az, J.~Lisenfeld, D.~Marcos,
J.~J.~Garc\'{\i}a-Ripoll, E.~Solano, C.~J.~P.~M.~Harmans, and
J.~E.~Mooij, Phys. Rev. Lett. 105, 237001
(2010).
\bibitem{PRL-105-196402-2010} Y.~Todorov, A.~M.~Andrews,
R.~Colombelli, \emph{et al.}, Phys. Rev. Lett. 105, 196402 (2010).
\bibitem{PRL-106-196405-2011} T.~Schwartz, J.~A.~Hutchison, C.~Genet, and
T.~W.~Ebbesen, Phys. Rev. Lett. 106, 196405 (2011).
\bibitem{Science-335-1323-2012} G.~Scalari, C.~Maissen,
D.~Tur\v{c}inkov\'{a}, \emph{et al.} Science 335, 1323 (2012).
\bibitem{PRL-108-163601-2012} A.~Crespi, S.~Longhi, and R.~Osellame,
Phys. Rev. Lett. 108, 163601 (2012).
\bibitem{PRB-86-045408-2012} S.~Hayashi, Y.~Ishigaki, and M.~Fujii, Phys.
Rev. B 86, 045408 (2012).

\bibitem{IEEE-51-89-1963} E.~T.~Jaynes and F.~W.~Cummings, Proc.
IEEE 51, 89 (1963); S.~B.~Zheng and G.~C.~Guo, Phys. Rev. Lett. 85,
2392 (2000).

\bibitem{NJP-13-073002-2011} X.~Cao, J.~Q.~You, H.~Zheng, and F.~Nori,
New J. Phys. 13, 073002 (2011).
\bibitem{PRL-109-193602-2012} A.~Ridolfo, M.~Leib, S.~Savasta, and M.~J.~Hartmann,
Phys. Rev. Lett. 109, 193602 (2012).
\bibitem{PRA-87-013826-2013} S.~Ashhab,
Phys. Rev. A 87, 013826 (2013).
\bibitem{PRA-59-4589-1999} H.~P.~Zheng, F.~C.~Lin, Y.~Z.~Wang, and
Y.~Segawa, Phys. Rev. A 59, 4589 (1999).
\bibitem{PRA-62-033807-2000} S.~B.~Zheng, X.~W.~Zhu, and M.~Feng,
Phys. Rev. A 62, 033807 (2000).
\bibitem{PRB-72-195410-2005} E.~K.~Irish, J.~Gea-Banacloche, I.~Martin, and K.~C.~Schwab, Phys.
Rev. B 72, 195410 (2005).
\bibitem{PRA-74-033811-2006} C.~Ciuti and I.~Carusotto, Phys.
Rev. A 74, 033811 (2006).
\bibitem{PRA-77-053808-2008} D.~Wang, T.~Hansson, {\AA}.~Larson, H.~O.~Karlsson, and J.~Larson, Phys.
Rev. A 77, 053808 (2008).
\bibitem{PRA-82-022119-2010} X.~F.~Cao, J.~Q.~You, H.~Zheng, A.~G.~Kofman, and F.~Nori, Phys.
Rev. A 82, 022119 (2010).
\bibitem{PRL-107-190402-2011} P.~Nataf and C.~Ciuti,
Phys. Rev. Lett. 107, 190402 (2011).
\bibitem{PRL-108-180401-2012} V.~V.~Albert,
Phys. Rev. Lett. 108, 180401 (2012).
\bibitem{PRA-87-022124-2013} F.~Altintas and R.~Eryigit, Phys.
Rev. A 87, 022124 (2013).
\bibitem{PRA-86-014303-2012} L.~H.~Du, X.~F.~Zhou, Z.~W.~Zhou, X.~Zhou, and G.~C.~Guo, Phys.
Rev. A 86, 014303 (2012).



\bibitem{JPA-29-4035-1996} I.~D.~Feranchuk, L.~I.~Komarov, and A.~P.~Ulyanenkov, J. Phys. A: Math. Gen. 29, 4035 (1996).
\bibitem{EPL-96-14003-2011} Q.~H.~Chen, T.~Liu, Y.~Y.~Zhang, and K.~L.~Wang, Eur. Phys.
Lett. 96, 14003 (2011).

\bibitem{PRA-81-042311-2010} S.~Ashhab and F.~Nori, Phys.
Rev. A 81, 042311 (2010).


\bibitem{RPB-40-11326-1989} H.~Chen, Y.~M.~Zhang, and X.~Wu, Phys. Rev. B 40,
11326 (1989).
\bibitem{PRB-42-6704-1990} J.~Stolze and L.~M\"{u}ller, Phys. Rev. B 42,
6704 (1990).
\bibitem{PRL-99-173601-2007} E.~K.~Irish, Phys. Rev. Lett. 99, 173601 (2007).
\bibitem{EPL-86-54003-2009} T.~Liu, K.~L.~Wang, and M.~Feng, Eur. Phys.
Lett. 86, 54003 (2009).
\bibitem{PRA-80-033846-2009} D.~Zueco, G.~M.~Reuther, S.~Kohler, and P.~H\"{a}nggi, Phys.
Rev. A 80, 033846 (2009).
\bibitem{PRL-105-263603-2010} J.~Casanova, G.~Romero, I.~Lizuain,
J.~J.~Garc\'{\i}a-Ripoll, and E.~Solano, Phys. Rev. Lett. 105,
263603 (2010).
\bibitem{PRA-82-025802-2010} M.~J.~Hwang and M.~S.~Choi, Phys.
Rev. A 82, 025802 (2010).
\bibitem{PRL-107-100401-2011} D.~Braak, Phys. Rev. Lett. 107,
100401 (2011).
\bibitem{EPJD-66-1-2012} J.~Song, Y.~Xia, X.~D.~Sun, Y.~Zhang, B.~Liu, and H.~S.~Song,
 Eur. Phys. J. D 66, 1 (2012).
\bibitem{PRA-86-015803-2012} L.~X.~Yu, S.~Q.~Zhu, Q.~F.~Liang, G.~Chen, and S.~T.~Jia, Phys.
Rev. A 86, 015803 (2012).
\bibitem{PRA-85-043815-2012} S.~Agarwal, S.~M.~H.~Rafsanjani, and J.~H.~Eberly, Phys.
Rev. A 85, 043815 (2012).
\bibitem{PRA-86-023822-2012} Q.~H.~Chen, C.~Wang, S.~He, T.~Liu, and K.~L.~Wang, Phys.
Rev. A 86, 023822 (2012).

\bibitem{EPJB-38-559-2004} H.~Zheng, Eur. Phys. J. B
38, 559 (2004).
\bibitem{PRB-75-054302-2007} Z.~G.~L\"{u} and H.~Zheng, Phys. Rev. B 75,
054302 (2007).
\bibitem{EPJD-59-473-2010} C.~J.~Gan and H.~Zheng, Eur. Phys. J. D
59, 473 (2010).


\bibitem{arXiv-1303-3367v2-2013} K.~M.~C.~Lee and
C.~K.~Law, Phys. Rev. A 88, 015802 (2013).
\bibitem{arXiv-1304-2529v1-2013} D.~Braak, arXiv:1304.2529v1 (2013).



\bibitem{PRA-68-012101-2003} X.~G.~Wang and B.~C.~Sanders, Phys.
Rev. A 68, 012101 (2003).
\bibitem{EPJD-18-385-2002} X.~Wang and K.~M{\o}lmer, Eur. Phys. J. D 18, 385 (2002).

\end{thebibliography}
\end{document}